\documentclass[journal=acsnano,manuscript=article]{achemso}
\setkeys{acs}{articletitle = true}
\usepackage[version=3]{mhchem} 
\usepackage[T1]{fontenc}       
\usepackage{geometry}
\usepackage{natbib}
\usepackage{float}
\usepackage{caption}
\usepackage{amsmath}
\usepackage{siunitx}
\usepackage{todonotes}
\usepackage[normalem]{ulem}
\usepackage{color}

\author{Patrick Figliozzi}
\affiliation{Department of Chemistry and James Franck Institute, The University of Chicago,  929 E. 57th Street, Chicago, Illinois 60637, USA}

\author{Curtis W. Peterson}
\affiliation{Department of Chemistry and James Franck Institute, The University of Chicago,  929 E. 57th Street, Chicago, Illinois 60637, USA}

\author{Stuart A. Rice}
\affiliation{Department of Chemistry and James Franck Institute, The University of Chicago,  929 E. 57th Street, Chicago, Illinois 60637, USA}

\author{Norbert F.~Scherer}
\affiliation{Department of Chemistry and James Franck Institute, The University of Chicago,  929 E. 57th Street, Chicago, Illinois 60637, USA}
\email{nfschere@uchicago.edu}

\title
{Direct Visualization of Barrier Crossing Dynamics in a Driven Optical Matter System}

\begin{document}




\begin{abstract}
	A major impediment to a more complete understanding of barrier crossing and other single-molecule processes is the inability to directly visualize the trajectories and dynamics of atoms and molecules in reactions. Rather, the kinetics are inferred from ensemble measurements or the position of a transducer (\textit{e.g.} an AFM cantilever) as a surrogate variable. Direct visualization is highly desirable. Here, we achieve the direct measurement of barrier crossing trajectories by using optical microscopy to observe position and orientation changes of pairs of Ag nanoparticles in an optical ring trap, \textit{i.e.} passing events. A two-step mechanism similar to a bimolecular exchange reaction is revealed by analysis that combines detailed knowledge of each trajectory, a statistically significant number of repetitions of the passing events, and the driving force-dependence of the process. We find that while the total event rate increases with driving force, this increase is only due to increased rate of encounters. There is no drive force-dependence on the rate of barrier crossing because the key motion for the process involves a radial motion of one particle as a thermal (random) fluctuation allowing the other to pass. This simple experiment can readily be extended to study more complex barrier crossing processes by replacing the spherical metal nanoparticles with anisotropic ones or through more intricate optical trapping potentials.
\end{abstract}

\section{Introduction}

Chemical and physical processes are commonly represented in terms of ensemble averages that provide a link between microscopic and macroscopic dynamics. While the microscopic details of a process may vary from one realization to another, one can obtain an ensemble averaged macroscopic description of the process in the form of a kinetic rate law\cite{berry2000physical}. These statistical interpretations of a process do not provide detailed descriptions of individual particle motion and various deviations from an averaged macroscopic mechanism. The development of new techniques that allow chemical and physical processes to be studied on an individual event or molecule basis, together with growing realization of the ubiquity and variety of important processes that are determined by single-molecule motion, have brought "single-molecule" measurements to the forefront of the physical sciences\cite{fleming1990,rief1997,lu1998,trepagnier2004,Neuman,Presse2013,Shen2014,Neupane2016}. The many repetitions of identical experiments that characterize single particle (\textit{e.g.} molecule) measurements replace ensemble averages with probability distributions and families of trajectories that can be used to link the single-molecule and macroscopic properties of a process and separate the common and the fluctuating contributions to the particle dynamics\cite{Floyd2010}.
\par
Microscopic visualization of particles in an optical trap and the consequences of their manipulation with external fields has had a large impact in single molecule biophysics\cite{Svoboda1994,moffitt2008}. Whereas most such studies use a typically micron-scale visualized particle (or AFM cantilever)\cite{butt2005} to report on or manipulate the molecule(s) it is attached to, nano- and meso-scale particles can be systems of investigation in and of themselves\cite{yan2014,yan2015,trepagnier2004,Rondin2017,sule2017,shao2018}. Both classes of experiments, \textit{i.e.} reporting on cognate molecules or the particle systems themselves, can be readily repeated under uniform conditions, allowing kinetic data to be extracted. The high level of spatial and temporal detail combined with the potential to obtain a statistically significant number of repetitions in optical trapping experiments makes them an ideal system in which to study the link between the microscopic and macroscopic (\textit{i.e.} kinetic) behavior of a system. 
\par
In the present paper, we study the physical passing of particles in an optical ring trap and do so for different driving forces. The motion of single Ag nanoparticles is measured by (darkfield) digital microscopy and precise tracking of each particle from frame to frame allows the process to be studied with a high level of microscopic detail. The large number of Ag nanoparticle trajectories allows obtaining a detailed kinetic description of the process. Our studies involve plasmonic nanoparticles that are confined to quasi-one-dimensional optical ring traps and subjected to a controlled driving force. These particles feel a variety of forces that can all be leveraged to change the energy landscapes and driving forces. The dynamics of particle passing, \textit{i.e.} a sign change in the orientation of a particle pair, are influenced by the combined effect of the electrodynamic forces confining the particles to the ring trap\cite{Dholakia2010}, the random thermal forces expressed as Brownian motion of the particles\cite{einstein1905}, and the electrodynamic driving force that propels the particles around the ring\cite{Roichman2008}.
\par
The present experiment involving the visualization of a driven optical matter system is designed to mimic the steps of a bimolecular reaction\cite{Ovchinnikov1978}. The highly detailed experimental data allowed us to recognize and validate a two-step mechanism analogous to an exchange reaction for the particle passing process involving an encounter complex and progressing through a transition state. The first step depends on the driving force in the ring, while the second step is a thermally activated process without this dependence. We created a stochastic microscopic model that reproduces ensemble-level measurements using input data from a large number of independent trajectories in order to describe the second step. The advantage offered by our system is that it allows full and explicit characterization of particle dynamics vs. the dynamics that are presumed to take place on the molecular size and timescales. 

\section{Experimental Results and Discussion}
\begin{figure}
	\includegraphics[width=0.75\columnwidth]{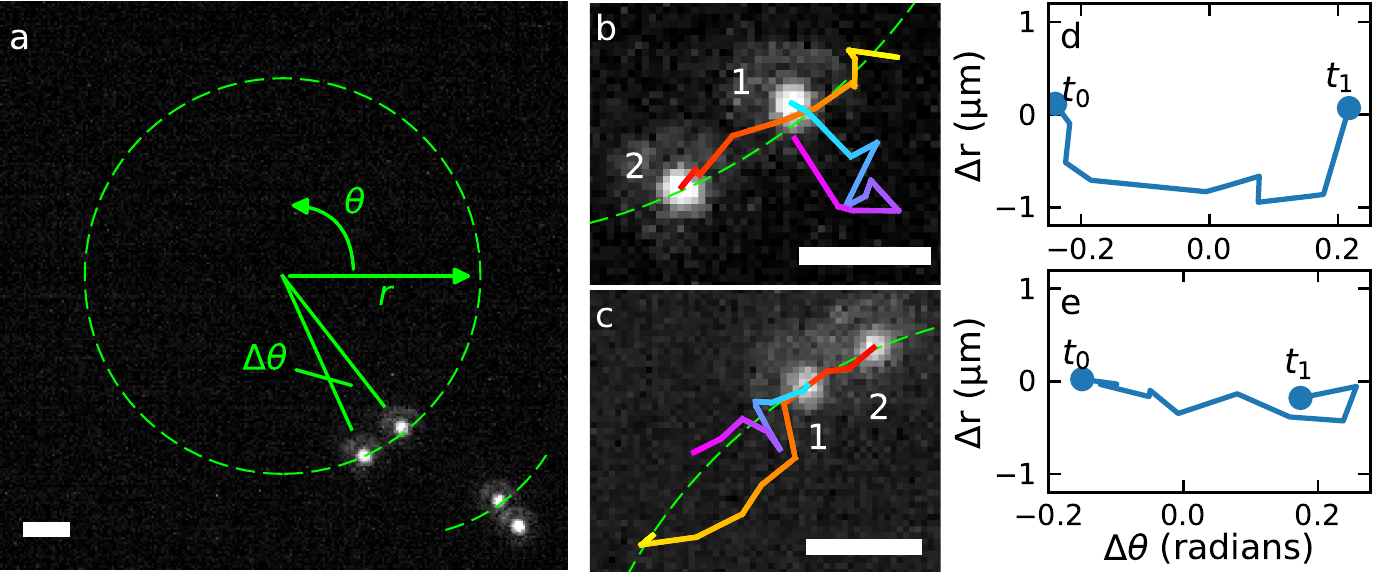}
	\caption{A pair of Ag nanoparticles in an optical ring trap and representative passing events that can occur. (a) Two Ag nanoparticles in a ring trap (dashed circle) with radius $r_{0} = 4.5 \mu m$. The inset shows a later time-step where one of the particles has fluctuated off the ring. The coordinates $r$ and $\theta$ are also shown in a. The driving force is in the counter-clockwise direction. (b,c) Two examples of passing events (each at $L = 5$ and $\approx 0.1 s$ in duration) with the changing color on the particle path representing time propagation. In (b) the leading particle fluctuates away from the radius $r_{0}$ of the ring trap, while in (c) the trailing particle fluctuates from the ring trap and simultaneously passes the leading particle. (d,e) Trajectories of the passing events shown in (b) and (c), respectively, in terms of relative coordinates $\Delta \theta$ and $\Delta r$.}
	\label{fig:Figure_1}
\end{figure}

\textbf{Ag nanoparticle trapping and passing} Ag nanoparticles were trapped and driven in a transverse plane over a glass coverslip using an optical trap ring  as described previously by Figliozzi \textit{et al.}\cite{Figliozzi2017}.
Briefly, an \SI{800}{\nano\meter} laser was reflected from a spatial light modulator (SLM) acquiring a suitable phase-encoded profile to create an optical ring trap when focused by a microscope objective (Olympus 60x water).
The power of the optical beam after the SLM and before the back aperture of the objective was \SI{40}{\milli\watt}.
A strong scattering force caused the \SI{150}{\nano\meter} diameter Ag nanoparticles to be held close to the glass surface balanced by electrostatic repulsion of the charged particles from the charged glass surface. An azimuthal phase gradient in the optical ring trap caused the nanoparticles to be driven around the trap along a quasi-one-dimensional path. (Figure~\ref{fig:Figure_1}a). The driving force in the optical ring trap was controlled by the topological charge $l$ (the number of $2\pi$ phase wrappings in one complete circuit around the ring) of the ring trap, which was varied from $l = 1$ to $l =5$ in the present experiment.

Figure~\ref{fig:Figure_1}a shows an image (raw data) of two Ag NPs in the trap. The arrow indicates their direction of directed motion. The laser power was lower in the present experiments compared to optimal trapping conditions in our previous study\cite{Figliozzi2017} to reduce the strength of the radial confinement of the Ag NPs. As a result, particles in the trap travel around the ring at a slower rate and have a wider radial distribution due to the diminished transverse intensity gradient force; they can undergo Brownian fluctuations in the radial direction and can pass each other due to radial position fluctuations, as shown in the inset to Figure~\ref{fig:Figure_1}a. The trajectories of the Ag nanoparticles in the optical ring trap are naturally described in a polar coordinate system, $r$ and $\theta$, as shown in Figure~\ref{fig:Figure_1}a.
The polar coordinates for each experiment were calculated by using a least squares routine to fit a circle of radius $r_{0}$ to the positions of all Ag NPs accumulated over a single experiment for a given value of $l$\cite{Allan2016}.

We define a passing event using a relative coordinate system, $\Delta r = r_{1} - r_{2}$ and $\Delta \theta = \theta_{1} - \theta_{2}$ where the subscripts $1$ and $2$ refer to the particles that are initially leading and trailing, respectively. A passing event occurs when there is a sign change in $\Delta \theta$. In general, a particle pair takes a random path through the two dimensional coordinate space ($\Delta r, \Delta \theta$) during such an event. Figure~\ref{fig:Figure_1}b shows the trajectories of both particles in a pair during a passing event. The chronological evolution of each particle's motion is encoded in color (red to yellow and magenta to blue for the trailing and leading particles, respectively). In this example the leading particle (in the direction of the applied driving force) fluctuates radially away from the mean radius $r_{0}$ of the ring trap while the trailing particle remains near $r_{0}$ and passes the lead particles driven by the applied optical force. A second passing event, along with individual particle motions, is shown in Figure~\ref{fig:Figure_1}c. Note that in this example the trailing particle passes around the leading particle.

The trajectories of the passing events shown in Figures~\ref{fig:Figure_1}b,c are shown in terms of ($\Delta r, \Delta \theta$) in Figures~\ref{fig:Figure_1}d and e, respectively.
The trajectories of passing events always start with $\Delta \theta < 0$ and progress to $\Delta \theta > 0$ because the relative coordinate system is designed with the leading particle at the origin.
The beginning and endpoints of the trajectory in ($\Delta r, \Delta \theta$) are indicated in Figure~\ref{fig:Figure_1}d and e by the time points $t_0$ and $t_1$, respectively.

\begin{figure}
	\includegraphics[width=.5\columnwidth]{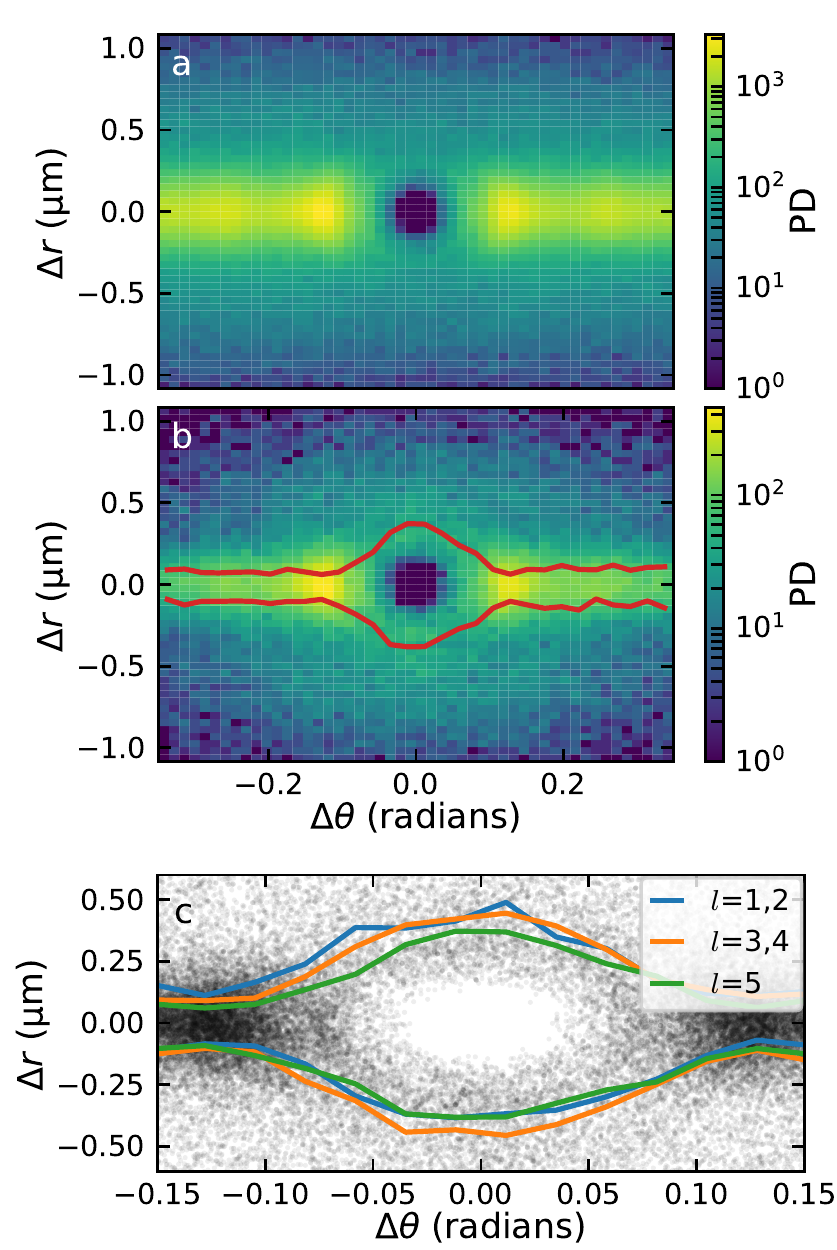}
	\caption{Probability density in relative coordinates $(\Delta \theta,\Delta r)$ and most probable paths for passing. (a) Total probability density over all experiments for all driving forces. (b) Conditional probability density in 30 frame windows centered on each passing events, over all experiments for all driving forces. (c) Scatter plot of points obtained from the same condition used in (b), with colored lines depicting the mean path of particles through $(\Delta \theta,\Delta r)$. Note that $\Delta \theta = 0.13rad = 600nm$ chord length so the regions of high point density correspond to optical binding.}
	\label{fig:Figure_2}
\end{figure}

If each passing event is defined as a particular trajectory through the two-dimensional coordinate space ($\Delta r, \Delta \theta$), the dynamics of the process will depend on the probability $P(\Delta r, \Delta \theta)$ of finding the system at a specific point in this space. Figure~\ref{fig:Figure_2}a shows this probability distribution for data aggregated over all experiments. We see that it is most likely to find $\Delta r$ near zero for $\Delta \theta > 0.1 rad$, which corresponds to a chord length of ~\SI{600}{\nano\meter}. This distance is associated with the expected separation for the electrodynamic interaction known as \textit{optical binding} at $\sqrt{\Delta r^{2} + (r\Delta \theta)^{2}} \approx \lambda_{incident}/n$,\cite{Burns1989,Dholakia2010,demergis2012} where $n$ is the index of refraction ($n = 1.33$ in water). However, it becomes extremely unlikely to find $\Delta r$ near zero for smaller values of $\Delta \theta$ due to electrostatic and electrodynamic repulsion between the charged Ag NPs\cite{sule2015}. Moreover, the particles never overlap in the images (videos) meaning they do not pass over each other in the axial direction of laser propagation. Therefore, for $\Delta \theta$ to be near zero at least one of the particles must be displaced off the ring (away from $r_{0}$), and therefore the passing process is 2-dimensional. 

Figure~\ref{fig:Figure_2}b shows a subset of the total probability density function (PDF) $P_{passing}(\Delta r, \Delta \theta)$ obtained by selecting only trajectories from a 30 frame window centered on each passing event. Applying this condition does not change the qualitative features of the PDF. The mean paths of the passing events (aggregated over all experiments) were separated depending on whether $\Delta r$ is positive or negative at $\Delta \theta =0$ are shown as red lines. These mean paths emphasize that the Ag NP passing process involves changes in both $\Delta r$ and $\Delta \theta$. Figure~\ref{fig:Figure_2}c shows a scatter plot of the points $(\Delta r, \Delta \theta)$ within a 30 frame window with the passing event at the center with the mean path through $(\Delta r, \Delta \theta)$ separated according to low (blue), medium (orange), and high (green) driving forces. The driving force appears to have, at most, a small effect on the mean path the system takes through $(\Delta r, \Delta \theta)$ during a passing event.

\newpage

\begin{figure}
	\includegraphics[width=.5\columnwidth]{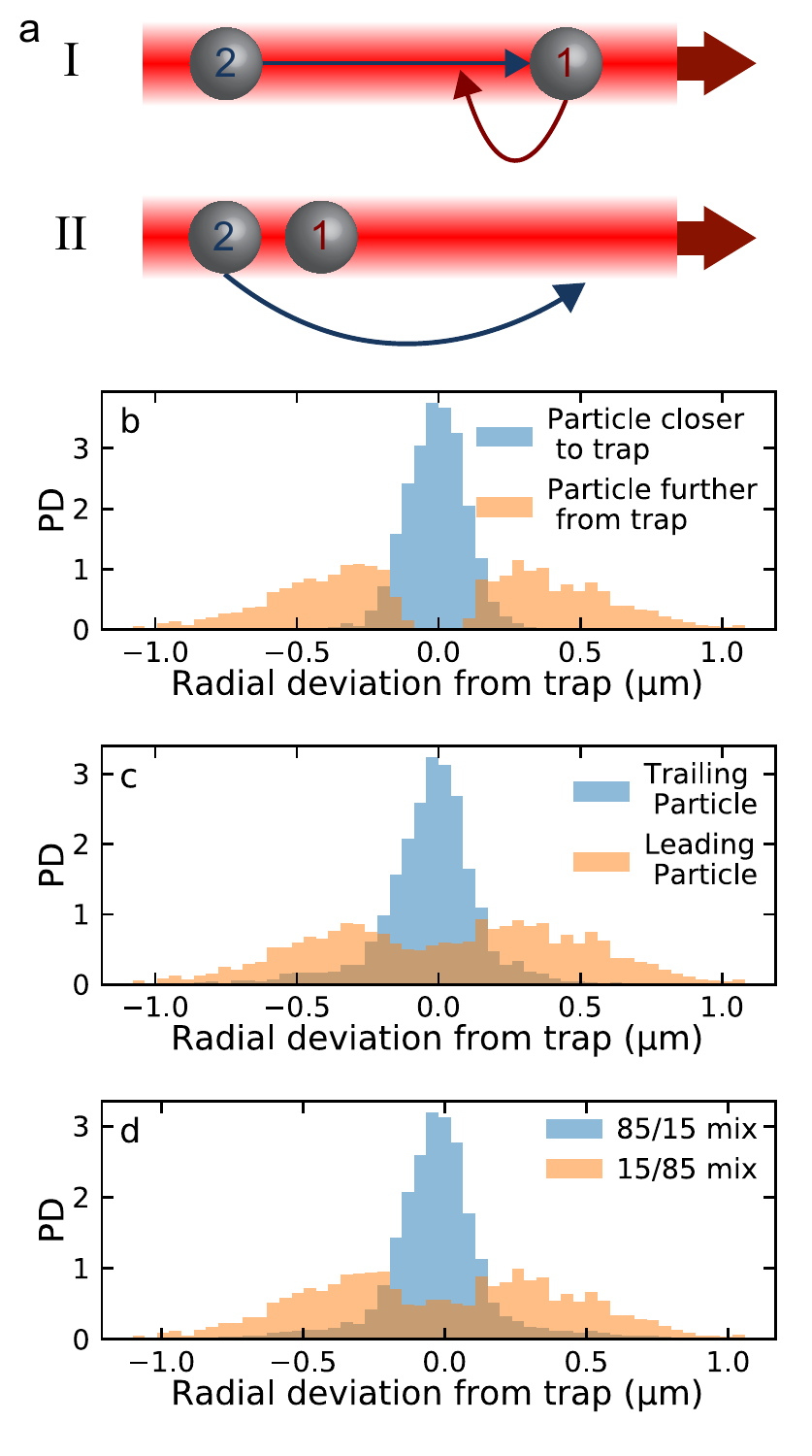}
	\caption{Schematic of two possible passing schemes and their related probability distributions. (a) Two possible schemes by which passing events take place. Scheme I is related to Figure~\ref{fig:Figure_1}b where the leading particle fluctuates away from the ring trap while the trailing particle moves past it. Scheme II is related to Figure~\ref{fig:Figure_1}c where the trailing particle fluctuates away from the trap and passes the leading particle. (b) PDFs of particle pair deviations from the ring trap during passing events for all values of $l$ for the particle closer to (blue) and further from (orange) $r_{0}$ at the time of passing(c) PDFs of particle pair deviations from the ring trap during passing events for all values of $l$ for the particle trailing (blue) and leading (orange) just before the event. (d) Remixing of the PDFs in (b) \textit{via} Equation~\ref{eq:remixing} with $C_{11}=C_{22}=0.85$ and $C_{12}=C_{21}=0.15$.}
	\label{fig:Figure_3}
\end{figure}

\textbf{Mechanism for passing.} So far we have discussed passing events in terms of quantities averaged over many trajectories. Because we have access to individual trajectories, however, it is possible to deduce a mechanism or mechanisms by which the passing occurs. Since $\Delta r$ must deviate from $0$ for a passing event to occur, it is important to determine the typical radial fluctuations of both particles involved in the event. Figure~\ref{fig:Figure_3}a show two likely mechanisms for the passing. In scheme I, the leading particle momentarily jumps away from the mean radius $r_{0}$ of the trap and the trailing particle then passes it. Conversely, in scheme II, the trailing particle jumps away from the mean radius $r_{0}$ of the trap while simultaneously passing the leading particle. In both of these schemes, only one particle fluctuates radially away from $r_{0}$. The trajectories shown in Figure~\ref{fig:Figure_1} b and c respectively reflect schemes I and II.

\par
Figure~\ref{fig:Figure_3}b shows the PDF of particle pair deviations from the ring trap at the time of passing combining events measured for all values of $l$ separated into two groups corresponding to the particle in the pair that is closer to $r_{0}$ and the particle that is further from $r_{0}$ at the time of the passing event. The particle closer to $r_{0}$ is approximately Gaussian distributed in $r$, while the particle further from the trap has no probability density at $r_{0}$. This result indicates that in the majority of passing events only one particle fluctuates radially while the other remains confined to the mean radius, $r_{0}$, of the ring. Figure~\ref{fig:Figure_3}c shows a splitting of the particle radial deviation PDFs when analyzed using a different condition: the PDFs were split according to whether a given particle was initially leading or trailing. This alternative condition results in slightly different PDFs compared to those shown in Figure~\ref{fig:Figure_3}a, which implies that either the leading or trailing particle can both be the one to fluctuate radially away from $r_{0}$, as depicted in schemes I and II in Figure~\ref{fig:Figure_3}a. 
\par
To estimate the prevalence of each mechanism, we re-mixed the PDFs in Figure~\ref{fig:Figure_3}b in different proportions according to the relationship

\begin{equation}
\begin{aligned}
P_{1}'=C_{11}P_{1} + C_{12}P_{2}
\\
P_{2}'=C_{21}P_{1} + C_{22}P_{2}
\end{aligned}
\label{eq:remixing}
\end{equation}

\noindent
where $P_{1}$ and $P_{2}$ are the PDFs of the particles closer to and further from $r_{0}$ in Figure~\ref{fig:Figure_3}b, respectively, while $P_{1}'$ and $P_{2}'$ are the PDFs for the trailing and leading particles. The $C_{ij}$ are the coefficients that determine the proportion of the mixing. Mechanistically, the diagonal elements in $C_{ij}$ correspond to scheme I, while the off-diagonal elements correspond to scheme II. Figure~\ref{fig:Figure_3}d shows the histograms obtained when $C_{11}=C_{22}=0.85$ and $C_{12}=C_{21}=0.15$, which gives the best match between Figure~\ref{fig:Figure_3}c and d. Therefore, the passing events occur $85\%$ of the time \textit{via} scheme I and 15$\%$ of the time \textit{via} scheme II . 

\begin{figure}
	\includegraphics[width=.5\columnwidth]{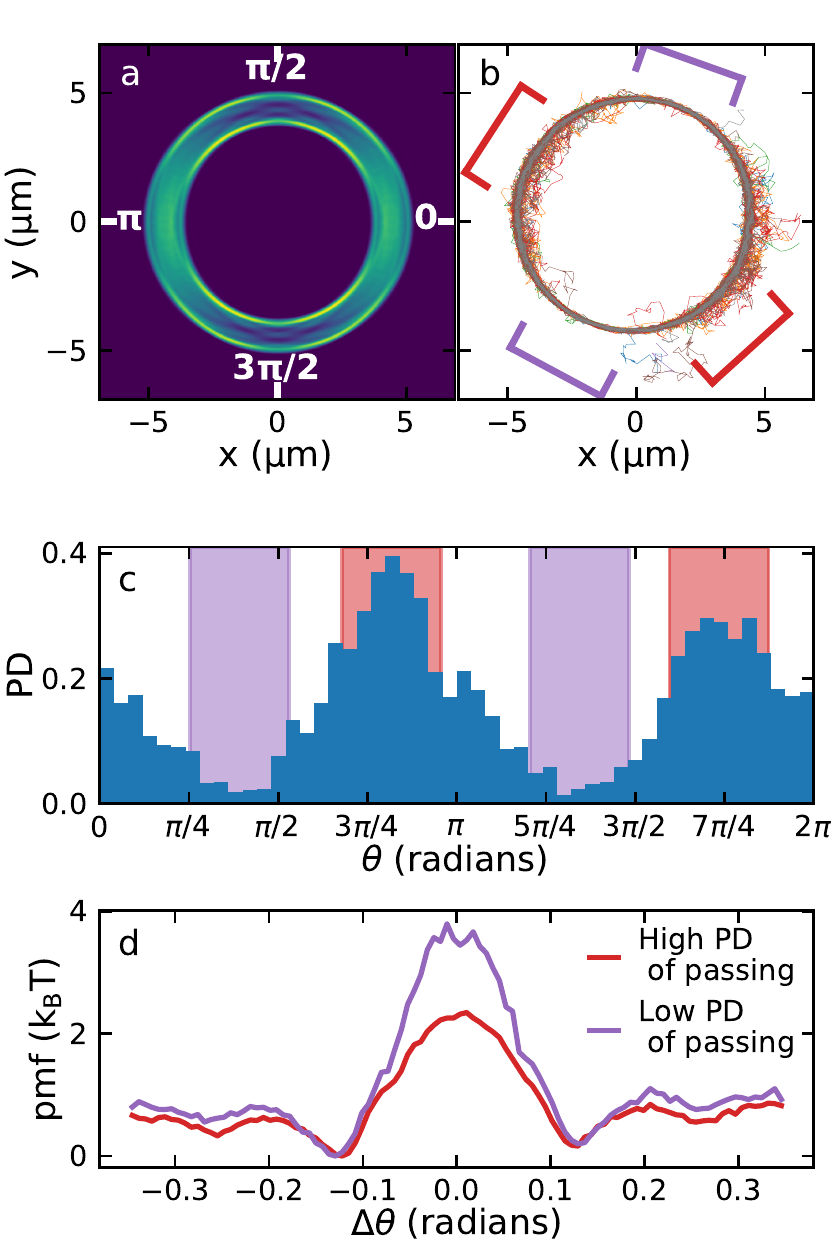}
	\caption{Factors that can affect the passing event location and barrier height changes for the passing process. (a) Simulated optical binding potential obtained by extending the pair-wise electrodynamic potential for a particle with fixed optical polarization to the pair on a circle of the same radius as the experimental ring trap, and weighting it by the probability of finding a particle at each $\theta$ position on the ring. (b) Superposition of single Ag NP trajectories for a representative experiment. Radial fluctuations away from $r_{0}$ are much more likely at $2\pi /3$ and $5\pi /3$ compared to $\pi /3$ and $4\pi /3$. The red and purple shading denotes regions of high and low passing event probability, respectively. (c) Probability density of passing events vs. angular position on the ring trap. (d) Potential of mean force in coordinate $\Delta \theta$ in areas of high (red) and low (purple) probability of passing event occurrence.}
	\label{fig:Figure_4}
\end{figure}

\textbf{Electrodynamic interactions and potentials of mean force.} We now turn to the question of how passing events depend on the electrodynamic interactions between particles, and the electrodynamic potential created by the ring trap. One important type of electrodynamic interaction between trapped particles is optical binding \cite{Dholakia2010,Burns1989}, which results from the incident electric field interfering with the scattered electric field from each particle. In our experiments the polarization state of the trapping beam is horizontal in the laboratory frame aligned along the 0 to $\pi$ coordinate of the ring shown in Figure~\ref{fig:Figure_1}a. The optical binding interaction is strongest between nanoparticles oriented perpendicular to the polarization. 
We simulated the optical binding potential in the optical ring trap by starting from the pair-wise electrodynamic potential\cite{yan2014} obtained from finite difference time domain (FDTD) simulations and extending these results around a circle of the same radius as the experimental ring trap and weighting it by the measured probability of finding a particle at each $\theta$ position on the ring (this probability is modulated by varying speed in different sections of the ring for linear polarization\cite{Figliozzi2017}). 
Figure~\ref{fig:Figure_4}a shows this estimated optical binding potential for a pair of Ag NPs around the ring trap. The optical binding interaction is most likely to stabilize a particle position away from the ring near $\pi /2$ and $3\pi /2$, and the interaction is symmetric about these points.

The actual tightness or looseness of single Ag NP confinement created by the optical ring trap in the radial direction is visualized in Figure~\ref{fig:Figure_4}b, which shows all \underline{single particle} trajectories in an experiment ($l=5$). It is apparent that deviations from $r_{0}$ are much more common near $2\pi /3$ and $5\pi /3$ compared to $\pi /3$ and $4\pi /3$. The reason for diminished confinement in these regions is a slight astigmatism introduced to the trap\cite{born2013,wang1980}. Therefore, we can ascertain whether passing behaves as a concerted rotation of the optically bound particle pair or as spontaneous radial fluctuations in the regions of reduced confinement.

\par

The distribution of passing events with respect to $\theta$ shown in Figure~\ref{fig:Figure_4}c indicates that passing events have maximum probability density near $2\pi /3$ and $5\pi /3$ and minimum probability density near $\pi /3$ and $4\pi /3$. This distribution is clearly dominated by the reduced radial confinement effect depicted in Figure~\ref{fig:Figure_4}b compared to the electrodynamic binding of Figure~\ref{fig:Figure_4}a. Therefore, somewhat surprisingly given the obvious presence of optical binding in Figure~\ref{fig:Figure_2}c, the single particle dynamics are much more important than interactions between particles with regard to the passing mechanism.

\par

The $\theta$ dependence of passing event probability reflects a barrier to the passing event whose height depends on the angular position of the particles in the ring. We can construct a potential of mean force (PMF) in the $\Delta \theta$ coordinate around a particular value of $\theta$, which we will denote $\theta_{0}$ by considering the conditional probability distribution $P(\Delta \theta|\theta \in [\theta_{0} - \delta,\theta_{0} + \delta])$, \textit{i.e.} the probability distribution of the angular separation $\Delta \theta$ given that the position of a particle pair $\theta$ is within some range $\pm \delta$ of $\theta_{0}$, the point of interest on the ring. Figure~\ref{fig:Figure_4}d shows the conditional PMF with $\theta_{0}$ at the centers of the red (high passing probability)  and purple (low passing probability) regions in Figure~\ref{fig:Figure_4}c. The increased rate of passing near $2\pi /3$ and $5\pi /3$ compared to $\pi /3$ and $4\pi /3$ corresponds to a barrier that is about 1.5 $k_{B}T$ lower in the regions of high passing probability compared to the low probability regions. Since this free energy landscape more closely resembles Figure~\ref{fig:Figure_4}b compared to Figure~\ref{fig:Figure_4}a it is consistent with the second step in the mechanism for passing depending primarily on the single-particle potential of the optical trap rather than on interactions between particles.

\newpage

\par
\noindent

\begin{figure}
	\includegraphics[width=.5\columnwidth]{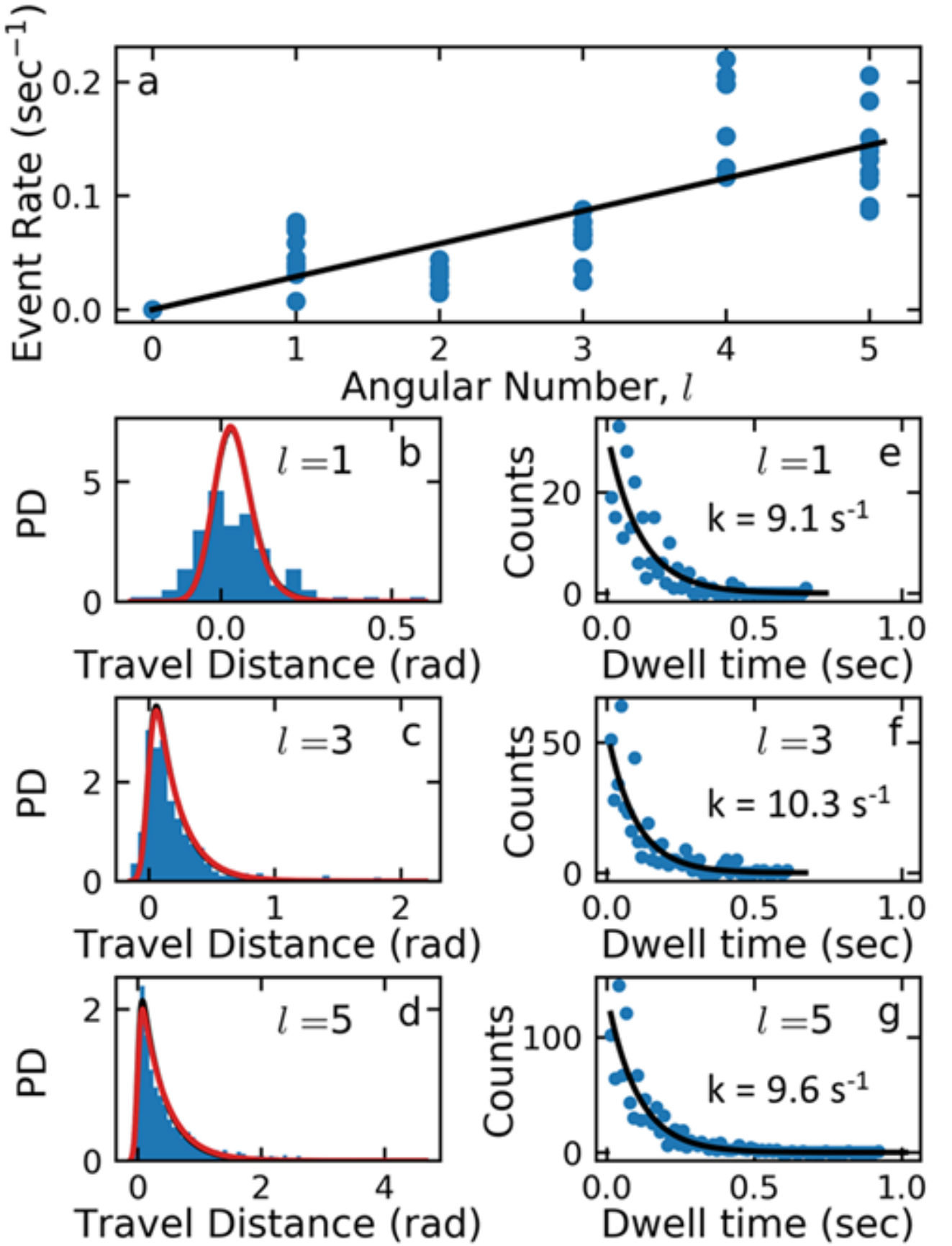}
	\caption{Kinetics of passing events and relation to a two-step stochastic mechanism. (a) Rate of passing events vs. driving force. (b-d) Distance traveled by a particle pair from formation to completion of the passing event. The solid red lines are theoretical histograms simulated \textit{via} equation~\ref{eq:traveldist}. (e-g) Distributions of event times from start to finish corresponding to the events in (b-d).}
	\label{fig:Figure_5}
\end{figure}

\textbf{Analogy to bimolecular mechanism analogy.} Given that the particles are being driven around the ring trap, elucidation of the mechanism for particle passing requires considering the effect of the electrodynamic driving force. Figure~\ref{fig:Figure_5}a shows the rate of events (per second) for driving forces increasing from $l=1$ to $l=5$. Since the data are collected from many experiments that have different numbers of particles n (\textit{i.e.} $n=2-6$ particles in the ring at the same time), the rates are normalized by the combinatorial number of possible particle pairs in a given experiment

\begin{equation}
\binom{n}{2}=\frac{n!}{2!(n-2)!}
\label{eq:paircombo}
\end{equation}

\noindent
where n is the number of particles in a given experiment. Figure~\ref{fig:Figure_5}a shows that the total event rate increases with with driving force. An increase in reaction rate with increasing driving force is predicted by both Arrhenius (or transition state) theory \cite{Bell1978} and Kramers theory\cite{Dudko2006}. However, since these theoretical descriptions were formulated for simple reaction mechanisms with single steps, it is necessary to establish a reaction mechanism to understand the increase in the "reaction" rate in our experiments. 

\par

To do this, we introduce a two step process analogous to a bimolecular exchange reaction

\begin{equation}
A+B {{\rightleftharpoons}} AB {{\rightleftharpoons}} AB^{\ddagger} {{\rightarrow}} B + A
\label{eq:twostep}
\end{equation}

\noindent
where $A+B$ are the two separated particles in their original (spatial) order, $AB$ is the particle pair once they are within a certain distance (\textit{i.e.} an optically bound pair that is analogous to an encounter complex), $AB^{\ddagger}$ is the structure at the transition state, and $B+A$ is the separated particle pair after the passing event with exchange of orientational order. In this mechanism, the total rate depends both on the formation of a particle pair (encounter complex), and an activated process to progress from the encounter complex to the reordered pair (product). The rate of the first step, forming the complex, should depend on the total number of particles, and we have accounted for it being proportional to the number of possible particle pairs by using equation~\ref{eq:paircombo}. However, it is not immediately obvious which step in the mechanism contains the driving force-dependence of the total rate. To address this, we consider the kinetics of the second step more closely. Figures~\ref{fig:Figure_5}b-d shows the distance traveled by a particle pair (within a certain threshold distance) from formation to completion of the passing event, and Figures~\ref{fig:Figure_5}e-g show the corresponding distribution of event times from pair formation to completion. The event time distributions show that once the particle pair is formed the process follows an exponential rate law that is not affected by driving force. 

\par
To confirm that once the encounter complex is formed the kinetics of the process no longer depend on driving force, we created a simple stochastic model for travel distance

\begin{equation}
d_{i}(v(l),D,\Delta t_{i})=N(v(l)\Delta t_{i},2D\Delta t_{i})=v(l)\Delta t_{i}+\sqrt{2D\Delta t_{i}}N(0,1)
\label{eq:traveldist}
\end{equation}

\noindent
where $d_{i}$ is the distance traveled in a particular realization of the process, $v$ is the driving force-dependent drift speed of a particle in the ring trap measured from experimental data, $D$ is the diffusion constant of a particle in the ring trap, $N(0,1)$ is a normal distribution with zero mean and variance of 1, and $\Delta t_{i}$ is the lifetime of the encounter complex, which is an exponentially distributed random variable. Figures~\ref{fig:Figure_5}e-g show exponential fits of experimentally measured lifetime distributions, which indicates the second step is a first order kinetic process described by

\begin{equation}
\frac{d  P(AB)}{dt} \propto e^{-kt} 
\label{eq:lifetime}
\end{equation}

\noindent
 where $k$ is the characteristic rate constant of the process and $P(AB)$ is the survival probability of the encounter complex. It is important to note that the distribution of $\Delta t_{i}$ does not appreciably change with drive force, and consequently $\Delta t_{i}$ were sampled from the same distribution obtained by averaging over results for all driving forces. We ran 10000 realizations of this process. The resulting PDFs, shown as red curves in Figures~\ref{fig:Figure_5}b-d, closely match our experimental data, validating our simple stochastic model.
 
 While the total rate of the passing event process increases with driving force, our simulations show that once the encounter complex, $AB$, is formed the driving force-dependence disappears. This implies that the driving force-dependence is contained completely in the first step in equation~\ref{eq:twostep}. In the language of chemical reactions, increasing the driving force has an effect analogous to increasing the frequency of encounters between reactants without affecting the energetics of the ensuing reaction. This is in contrast to theories commonly employed to predict the effect of a driving force on the rate of a reaction, where the driving force induces a tilt in the free energy surface that lowers the effective barrier of activation of the reaction\cite{Bell1978,Dudko2006}.

The mechanism of our Ag NP passing event process has now fully taken shape. In the first step two particles must approach each other to form an encounter complex. This encounter complex is at an optical binding separation of a particle pair. In fact, the dense collection of points in Figure~\ref{fig:Figure_2}c at $\Delta \theta = -0.12rad$ reflects this initial complex. We found the rate of formation of the encounter complex to depend both on the number of particles (through equation~\ref{eq:paircombo}) and the drive force  at as seen in Figure~\ref{fig:Figure_5}a. Once the optically bound encounter complex is formed, completion of the subsequent activated process obeys an exponential rate law. We concluded from Figure~\ref{fig:Figure_3} that typically only one particle fluctuates radially away from the ring trap, and most ($85 \%$) of the time the front particle is the one which undergoes this fluctuation. From the propensity for only a single particle to fluctuate radially away from $r_{0}$ and the kinetic data in Figure~\ref{fig:Figure_5}, it is apparent that the rate of the first order kinetic process is simply due to thermal forces pushing one of the particles out of the trap, as suggested in the schemes of Figure~\ref{fig:Figure_3}.

\par

\begin{figure}
	\includegraphics[width=.5\columnwidth]{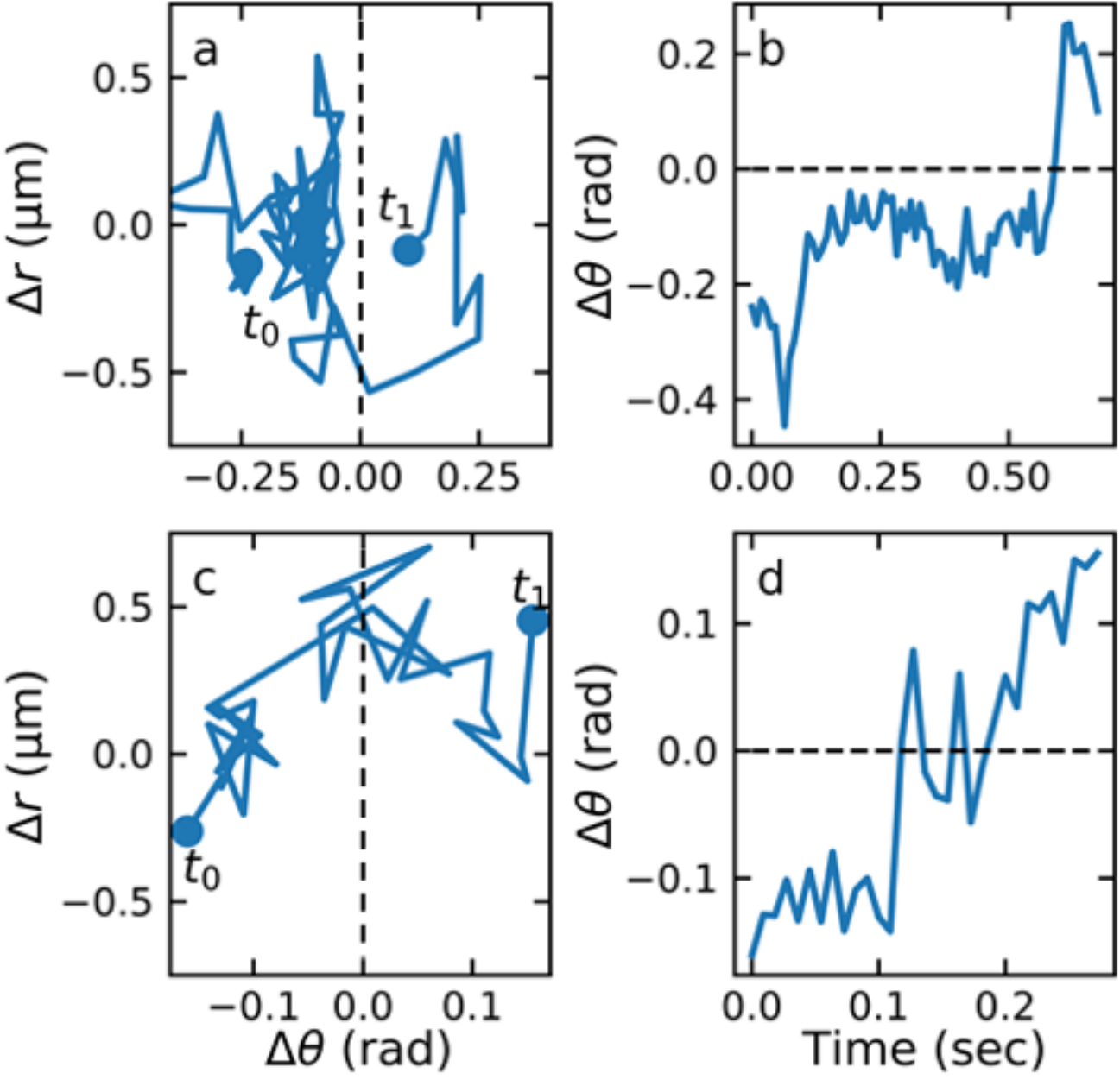}
	\caption{Barrier recrossing from detailed trajectory information available in optical trapping experiments. (a,c) Two trajectories that show significant waiting times in coordinates ($\Delta \theta$,$\Delta r$). (b,d) Trajectories from (a,c) in coordinates (t,$\Delta \theta$). Barrier recrossing about $\Delta\theta = 0$ is evident in (d).}
	\label{fig:Figure_6}
\end{figure}

\textbf{Barrier Crossing and Recrossing.} High time and spatial resolution in optical trapping experiments allows for determination of detailed trajectories through a barrier region, as seen in Figure~\ref{fig:Figure_6} a and b. While most passing event trajectories resemble those shown in Figure~\ref{fig:Figure_1} and Figure~\ref{fig:Figure_6}a,b, that is, single barrier crossing events, we also observe barrier recrossing. Figure~\ref{fig:Figure_6}c,d shows a passing event that involves multiple crossings before the process is complete. A number of trajectories exhibit such barrier re-crossing, implying that a more accurate analysis of this electrodynamically driven nanoparticle system should include a correction to transition state theory. 

\section{Conclusion}
Transition paths in thermally activated processes such as protein and DNA folding have only recently been related\cite{Neupane2016}, but the reaction coordinate was inferred from the position of beads connected to the molecule of interest using a handle (\textit{e.g.} ds-DNA). The response of both of these extraneous portions of the the experimental system are convoluted with the molecular signal of interest, complicating the experimental analysis\cite{Cossio2015}. In experiments where the system of interest is directly observed, dynamics and the reaction coordinate are also directly determined, eliminating these complications. Optical trapping experiments have the potential to explore questions regarding non-equilibrium transport at a single-particle level due to their ability to shape both conservative and non-conservative force fields.
\par
We studied individual passing events of pairs of Ag nanoparticles in an optical ring trap with a controlled adjustable driving force. Our detailed and precisely localized trajectory data measured over many realizations of this process along with stochastic model simulations allowed identification of a detailed mechanism casting this problem in close analogy with bimolecular exchange reactions in solution. The passing event process is analogous to a two-step bimolecular reaction with an encounter complex followed by barrier crossing as described by equation~\ref{eq:twostep} and the rate of passing events increases with driving force. Our detailed trajectories reveal a two step mechanism where the driving force increases the rate of the first step, while the second step is independent of driving force. Surprisingly, the second step is thermally activated barrier crossing of the encounter complex formed in the first step. The exponentially distributed survival probability of the encounter complex implies that the second step is a first order kinetic process, \textit{i.e.} there is a constant probability density at any given time that the reaction will progress to completion, so the two-step characteristic of the passing process is crucial to the explanation of this type of driving force-dependence. Furthermore, the decay rate of this survival probability does not depend on the azimuthally directed driving force, which suggests that the reaction coordinate for this step lies significantly in the radial ($\Delta r$) direction that is orthogonal to the $\theta$ and $\Delta\theta$ aspect of passing. Finally, the level of detail available in nanoparticle visualization experiments allowed direct observation of barrier recrossing. However, we do not treat this phenomenon in detail in this paper. 

The present paper is the first report of this new approach to study barrier crossing phenomena. Many variations and novel situations are envisioned for future studies. Optical traps can be shaped with high precision to design conservative and nonconservative forces, and strong inter-particle forces related to optical binding can be utilized to study the effects of interaction in these potentials. Therefore, experiments can be designed to extend our approach to other chemical and physical processes by tailoring specific forces and interactions to reflect the behavior in an analogous system or to examine idealized theoretical scenarios.

\section{Methods}

\textbf{Experimental} The experiments were preformed with 150nm diameter Ag nanoparticles held and driven in an optical ring vortex as previously described\cite{Yan2013,Figliozzi2017}. 
The \SI{800}{\nano\meter} beam from a Ti-Sapphire laser is phase modulated with a spatial light modulator (SLM; Hamamatsu X10468-02) to produce the optical ring vortex \cite{Roichman2007a,Yan2013}. The experiments used $\sim$\SI{45}{\milli\watt} beam power going into the back aperture of the microscope objective.
Citrate capped \SI{150}{\nano\meter} Ag nanoparticles (NanoComposix) are diluted 200x and placed into a sample chamber described previously \cite{Figliozzi2017}. 
The scattering force of the laser applied to the nanoparticles pushes them very close to the glass-water interface of the top coverslip of the fluid well.
The nanoparticles are held in one plane perpendicular to the optical axis due to a balance of the scattering force and the electrostatic repulsion the particles have with the electrically charged glass surface \cite{Figliozzi2017}.
The Ag nanoparticles are trapped and driven around the optical ring with a drive force determined by the number of azimuthal phase wrappings, $l$, applied in the phase modulation pattern on the SLM.
The motion of the Ag nanoparticles is visualized \textit{via} darkfield microscopy and captured with a sCMOS camera (Andor Neo) at 110 frames per second.
A variety of different experiments were performed at different $l$'s with each one consisting of \SIrange[range-phrase=--]{45}{90}{\second} (\num[group-separator={,}]{5000} to \num[group-separator={,}]{10000} frames) of video.
In order to resolve distinct particle shapes without blurring or distortion a camera exposure of \SIrange{2e-3}{6e-4}{\second} was used when capturing video.

\textbf{Particle Tracking} Particle trajectories were extracted from the video data using the Python particle tracking software package TrackPy \cite{Allan2016}. 
A cluster tracking algorithm in TrackPy is used to accurately track the nanoparticles even when two or more nanoparticles become part of a cluster \cite{VanderWel2017}. 
The optimal parameters for each experiment were determined by hand and were set so that the number of particles identified in each frame is consistent with the number of particles in the experiment.
Additionally, frames where the focus of the image drifted were removed from particle tracking as the particle tracking algorithm would find false positives in the de-focused image of the particles.
However, this method of particle localization uses the center-of-mass method which can lead to significant errors especially when particles come in close proximity, and the SPIFF algorithm was used to alleviate these errors\cite{Burov2016,yifat2017analysis}.
A refinement algorithm was used that improves the accuracy of the positions of the particles by performing a non-linear least-squares (NLLS) fit of a Gaussian function to each distribution of pixel intensities for each nanoparticle. This allows extracting the particle positions with much greater accuracy especially in the case of overlapping features.

\textbf{Author Contributions}
P.F. and C.P. contributed equally to this work. All authors participated in the writing of the manuscript.

\textbf{Notes}
The authors declare no competing financial interests.

\begin{acknowledgement}
	The authors acknowledge support from the Vannevar Bush Faculty Fellowship program sponsored by the Basic Research Office of the Assistant Secretary of Defense for Research and Engineering and funded by the Office of Naval Research through grant N00014-16-1-2502.
	\noindent
	We thank the University of Chicago MRSEC (NSF DMR-1420709) for partial support
\end{acknowledgement}

	%
	%
	%

\bibliography{bibliography}

\end{document}